\documentclass{article}
\usepackage{amsmath, amssymb, ulem}
\usepackage[dvips]{graphicx}
\usepackage[top=30truemm,bottom=30truemm,left=30truemm,right=30truemm]{geometry}
\numberwithin{equation}{section}

\def\lessim{\lower0.6ex\hbox{$\,$\vbox{\offinterlineskip\hbox{$<$}\vskip1pt\hbox{$\sim$}}$\,$}}
\def\grtsim{\lower0.6ex\hbox{$\,$\vbox{\offinterlineskip\hbox{$>$}\vskip1pt\hbox{$\sim$}}$\,$}}

\begin{document}

\title{Domain wall in $f(R)$ gravity}
\author{Natsuki Watanabe  \\
	\phantom{}  \\
	{\small {\it Department of Physics, Tokyo Metropolitan University,}}  \\
	{\small {\it Hachioji-shi, Tokyo 192-0397, Japan}}  \\
	{\small {\it watanabe-natsuki1@tmu.ac.jp}}
	}
\date{\empty}

\maketitle

\begin{abstract}
	The gravitational impact inside and outside of a domain wall is studied in the context of $f(R)$ gravity theory.
	The function $f(R)$ is found which satisfies the stability conditions.
	Our results imply that the domain wall may cause topological inflation.
\end{abstract}

\section{Introduction}  \label{intro}

When a symmetry is spontaneously broken by a phase transition, topological defects are produced
(see e.g., \cite{Durrer1999} and \cite{Brandenberger1998} for a review).
The generated defects are different depending upon the kind of broken symmetry.
For example, in the case when a discrete symmetry $Z_{2}$ breaks, a two dimensional planar defect is produced.
Such defects are called domain walls (DW).
If the spherical symmetry $O(3)$ is spontaneously broken to a rotational symmetry $U(1)$,
then a point-like (zero-dimensional) defect arises. It is a monopole.
As the other defects, cosmic strings and textures are also known.
They are the one- and three-dimensional topological defects, respectively.

It is assumed in the current cosmology that topological defects could be generated when the inflation occurred.
However, no topological defects have been discovered until now.
If there are some defects somewhere in our universe, their density should be very small.
Topological defects may also be the origin of the dark matter,
the dark energy and the large-scale structure of our universe.
Therefore, studies of topological defects are useful  to understand the history of our universe.
Because of that reasoning, topological defects were extensively studied in the past ---
see e.g., ref. \cite{Kibble1976} for the study of creation and evolution of topological defects,
and refs. \cite{Vilenkin1981} and \cite{Harari1990} about the gravitational fields originated from the topological defects. 
In addition, an inflationary model by using topological defects (called the topological inflation) was proposed
in \cite{Linde1994a} and \cite{Vilenkin1994}.
It is similar to the chaotic inflation \cite{LInde1983},
but the universe created by a topological defect continues to expand exponentially.
Applications of DW to the dark energy were proposed in \cite{Friedland2003}.

More recently, series of models describing inflation or dark energy were proposed by many authors.
Usually, an explanation of accelerated expansion needs assumes a matter with a negative pressure.
The slow-roll inflation (\cite{Albrecht:1982wi} and \cite{Linde:1981mu})
assumes the existence of a new scalar field called the inflaton.
There are also the theories describing the accelerated expansion of the universe by modifying the geometry of the universe.
They are called the modified gravity theories.
The simplest example  is the $f(R)$ gravity theory.

The $f(R)$ gravity is defined by replacing the $R$ term in the Einstein-Hilbert action with $f(R)$,
where $R$ is the Ricci scalar and $f(R)$ is an arbitrary function in terms of $R$.
At present, there are many inflationary and dark energy models by applying in $f(R)$ gravity.
The Starobinsky model \cite{Starobinsky1983} is well known as an inflationary model, and is still viable.
A well known  dark energy model is an {\it ad-hoc} model proposed by Appleby and Battye \cite{Appleby2007}
\footnote{
	We studied applications of Starobinsky and Appleby-Battye models
	in refs. \cite{Kaneda2010a} and \cite{Ketov2012g} respectively.
	}.
In addition to those models many other models were proposed.
There are many reviews by the different authors (see e.g., refs. \cite{Felice2010} and \cite{Sotiriou2010}).

Several studies applying the $f(R)$ gravity to monopoles and DW were also suggested.
In the context of General Relativity,
monopoles and DW can be represented by a spherically and a plane symmetric metric.
The de Sitter-Schwarzschild solution \cite{McVittie1933}
is known as the exact static spherically symmetric solution in General Relativity.
On the one hand,
Multam\"{a}ki \cite{Multamaki2006} found that the de Sitter-Schwarzschild solution is also valid in  the $f(R)$ gravity.
Sebastiani \cite{Sebastiani2011a} found an exact solution different from the de Sitter-Schwarzschild solution.
On the other hand, as a plane symmetric solution in General Relativity,
the Taub's solution \cite{Taub2004} is to be noticed.
In the context of the $f(R)$ gravity,
Sharif and Shamir \cite{Sharif2010} derived three constant-curvature solutions
which correspond to the Taub's, anti-de Sitter and self-similar solutions, respectively.
In addition, the gravitational fields of monopoles,
cosmic strings and DW in the Starobinsky model were studied in \cite{Audretsch1993}.
Most recently, Caram\^{e}s et al. \cite{Carames2011} calculated the gravitational field of a monopole as the $f(R)$ function.
The gravitational field created by monopoles was suggested by Barriola and Vilenkin \cite{Barriola1989}.
Caram\^{e}s et al. derived the corresponding $f(R)$ function in the case considered in  \cite{Barriola1989}.
Next the natural question arises: how about DW?
This paper is devoted to a  calculation of the corresponding $f(R)$ function in the case of DW.

Most of the existing studies address DW under the assumption that the thickness of DW is negligible for simplicity.
However, we consider here a thick DW and calculate the $f(R)$ function outside and inside of DW.
We employ the plane symmetric metric of Taub \cite{Taub2004}.
Widrow \cite{Widrow1989} studied the gravitational field of a thick DW in the context of General Relativity
and calculated its gravitational field outside and inside the DW by using certain approximation.
We essentially employ the same approximation \cite{Widrow1989} though in the context of $f(R)$ gravity.

This paper is organized as follows.
To begin with, we give an outline about DW in General Relativity in Sec.~\ref{generalDW}.
In Sec.~\ref{formalism} we briefly review the metric formalism of the $f(R)$ gravity theory.
The equations of motion are derived in Sec.~\ref{equations}. Our main part is given by Sec.~\ref{DWinf(R)}.
Finally, Sec.~\ref{Conc} is conclusion.
Throughout this paper we use the natural units, $c = \hbar = 1$,
and the space-time signature $\eta_{\mu \nu} = (1, - 1, - 1, - 1)$.

\section{DW in General Relativity}  \label{generalDW}

We recall how DW are described in General Relativity.
As was mentioned in the Introduction, a DW is related to a discrete symmetry breaking.
Such symmetry, for example, is present in the well-known Higgs-type potential.
The corresponding Lagrangian is given by 
\begin{equation}  \label{DWLag}
	\mathcal{L} = \frac{1}{2} g^{\mu \nu} \partial_{\mu} \phi \partial_{\nu} \phi - V(\phi),
\end{equation}
where $\phi$ is a scalar field and the potential $V(\phi)$ is 
\begin{equation}  \label{DWpot}
	V(\phi) = \frac{\lambda}{4} (\phi^{2} - \eta^{2})^{2}.
\end{equation}
In flat space-time $g_{\mu \nu} = \eta_{\mu \nu}$, the classical DW solution for the Lagrangian \eqref{DWLag} is given by
\begin{equation}  \label{phi}
	\phi_{0} = \eta \tanh \left( \frac{z}{\delta_{0}} \right),
\end{equation}
where $\delta_{0}$
\begin{equation}  \label{thick}
	\delta_{0} = \frac{1}{\eta} \sqrt{\frac{2}{\lambda}},
\end{equation}
measures the thickness of DW.
The subscript ``0'' of the scalar field $\phi_{0}$ and the thickness $\delta_{0}$ indicate those quantities ``in flat space''.
As regards the Lagrangian above, its relevant equation of motion for the DW case takes the form
\begin{equation}  \label{flateom}
	\phi_{0}'' = \frac{\partial V(\phi)}{\partial \phi},
\end{equation}
where the primes denote the derivatives with respect to $z$.
Integrating this equation once yields
\begin{equation}  \label{flateomint}
	\frac{1}{2} \phi_{0}'^{2} = V(\phi_{0}).
\end{equation}

The metric representing DW is plane symmetric \cite{Taub2004}.
Assuming that  the DW lies on $xy$ plane, the most general metric compatible with it is given by
\begin{equation}  \label{psmet}
	ds^{2} = A(t, z) (dt^{2} - dz^{2}) - B(t, z) (dx^{2} + dy^{2}).
\end{equation}
In the case of a static DW, the $A$ and $B$ take the form \cite{Taub2004},
\begin{equation}  \label{stmet}
	A = \frac{1}{\sqrt{1 + K z}},
	\quad
	B = 1 + K z.
\end{equation}

Actually, the space outside of a DW without thickness is flat except for $z = 0$ \cite{Taub2004}.
Indeed, the Ricci tensor to be derived from eq.~\eqref{stmet} vanishes.
It is also worth noticing that, if DW has no thickness, it cannot be static,
because such DW will collapse and finally become a singularity, say, a black hole.
However, if DW has a thickness, a static DW is allowed \cite{Widrow1989}.

We consider the case of a plane symmetric metric \eqref{psmet}.
In this section we gave the equation of motion in flat space \eqref{flateom}.
It changes slightly with the metric \eqref{psmet}, as we are going to show in the next sections.
We also compute all components of the energy-momentum tensor
and of the Ricci tensor corresponding to the plane symmetric metric.
However, next we briefly review the $f(R)$ gravity theory.

\section{$f(R)$ gravity}  \label{formalism}

As was mentioned in the Introduction, the $f(R)$ gravity theory is the simplest modified gravity theory.
It is the theory in which the $R$-term in the Einstein-Hilbert action
is replaced with an arbitrary function $f(R)$ in terms of the Ricci scalar $R$.
The action is given by
\begin{equation}  \label{act}
	S_{f} = - \frac{1}{2 \kappa^{2}} \int d^{4}x \sqrt{- g}~ f(R) + S_{\rm m},
\end{equation}
where $\kappa$ is the gravitational coupling constant, $\kappa^{2} = 8 \pi G_{\rm N}$,
$G_{\rm N}$ is the Newton's constant and $S_{\rm m}$ is the matter part.
In the Einstein General Relativity one gets the Einstein equations as the gravitational equations of motion.
In the $f(R)$ gravity the equations of motion are more complicated,
\begin{equation}  \label{freom}
	f_{R}(R) R_{\mu \nu} - \frac{1}{2} f(R) g_{\mu \nu}
	  + (g_{\mu \nu} \Box - \nabla_{\mu} \nabla_{\nu}) f_{R}(R)
		= \kappa^{2} T_{\mu \nu},
\end{equation}
where $f_{R}$ denotes the derivative with respect to $R$, 
\begin{equation}
	f_{R} = \frac{d f(R)}{d R},
\end{equation}
and $T_{\mu \nu}$ is the energy-momentum tensor,
\begin{equation}  \label{emten}
	T_{\mu \nu} = \frac{2}{\sqrt{- g}} \frac{\delta S_{\rm m}}{\delta g^{\mu \nu}}.
\end{equation}
Taking the trace of eq.~\eqref{freom} we have
\begin{equation}  \label{trace}
	f_{R}(R) R - 2 f(R) + 3 \Box f_{R}(R) = \kappa^{2} T.
\end{equation}
In this expression General Relativity corresponds to $f(R) = R$ and $f_{R}(R) = 1$.
Then eq.~\eqref{trace} becomes $R = - \kappa^{2} T$, and $R$ is directly related to the matter $T$.
In the context of the $f(R)$ gravity, $f_{R}(R) \neq {\rm const.}$, so $\Box f_{R}(R) \neq 0$.
It means that the new propagating scalar degree of freedom $\omega \equiv f_{R}(R)$ is present.
The dynamics of this scalar field (often called ``scalaron'') is described by eq.~\eqref{trace}.

From the expression of eq.~\eqref{trace}, we can find that $f(R)$ is represented by $f_{R}(R)$,
\begin{equation}  \label{ftofR}
	f(R) = \frac{1}{2} \left[ f_{R}(R) R + 3 \Box f_{R}(R) - \kappa^{2} T \right].
\end{equation}
Substituting it into eq.~\eqref{freom}, we find
\begin{equation}
	f_{R}(R) R_{\mu \nu} - \nabla_{\mu} \nabla_{\nu} f_{R}(R) - \kappa^{2} T_{\mu \nu}
		= \frac{g_{\mu \nu}}{4} \left[ f_{R}(R) - \Box f_{R}(R) - \kappa^{2} T \right].
\end{equation}
The right hand side of this equation does not depend upon any indices.
So, defining the quantity
\begin{equation}
	C_{\mu}
		= \frac{f_{R}(R) R_{\mu \mu} - \nabla_{\mu} \nabla_{\mu} f_{R}(R)
			- \kappa^{2} T_{\mu \mu}}{g_{\mu \mu}},
\end{equation}
with fixed indices, we can obtain the identity
\begin{equation}  \label{identity}
	C_{\mu} - C_{\nu} = 0.
\end{equation}
This identity corresponds to the gravitational field equations.
We will use it in the following sections.

The $f(R)$ function is also subject to some stability constraints.
They are called ``the stability conditions,'' and are given by
\begin{equation}  \label{stab}
	f_{R}(R) > 0,
	\quad
	f_{R R}(R) > 0,
\end{equation}
in our notation.
The former is the classical stability condition, which means that graviton is not a ghost.
In addition, the latter is the quantum stability condition, which means that scalaron is not a tachyon.
So, both together they  guarantee that the system does not get an instability \cite{Faraoni2007}.
We take them into account when we calculate the $f(R)$ function below.

\section{Equations of motion of DW}  \label{equations}

We derive the equations of motion of DW and the gravitational field equations
in the plane symmetric space-time in this section.
The metric of this space-time is given by eq.~\eqref{psmet}.
We suppose the DW to be static, in order to simplify our calculations, so that $A = A(z)$ and $B = B(z)$.
In the context of General Relativity,
a derivation of the equations with a plane symmetric metric was done by Widrow \cite{Widrow1989}.
He studied the gravitational effects inside and outside of thick and non-static DW.
The static plane symmetric solution in the $f(R)$ gravity was considered  by Sharif and Shamir \cite{Sharif2010}.

We start with the equations of motion.
It can be easily calculated from the Lagrangian \eqref{DWLag},
\begin{align}
	\phi'' + \frac{B'}{B} \phi' - A \frac{\partial V(\phi)}{\partial \phi} = 0,  \label{originaleom}
\end{align}
where we used
\begin{equation}
	g^{\mu \nu} \partial_{\mu} \partial_{\nu} \phi
		= \frac{1}{\sqrt{- g}} \partial_{\mu} (\sqrt{- g} g^{\mu \nu} \partial_{\nu} \phi).
\end{equation}

Next, the energy-momentum tensor is defined by
\begin{equation}
	T_{\mu \nu} = \partial_{\mu} \phi \partial_{\nu} \phi - g_{\mu \nu} \mathcal{L}.
\end{equation}
In fact, most of its components vanish.
The  non-vanishing components of the energy momentum tensor are
\begin{align}
	&
	T^{0}{}_{0} = T^{1}{}_{1} = T^{2}{}_{2} = \frac{1}{2} \phi'^{2} + V(\phi),  \label{T00} \\
	&
	T^{3}{}_{3} = - \frac{1}{2} \phi'^{2} - V(\phi).  \label{T33}
\end{align}
The $(0, 0)$-component $T_{0 0}$, gives the surface energy density of DW,
\begin{equation}  \label{sdef}
	\sigma = \int^{\infty}_{0} dz T_{0 0}.
\end{equation}
In flat space it becomes
\begin{equation}  \label{sflat}
	\sigma = \frac{2 \sqrt{2}}{3} \sqrt{\lambda} \eta^{3}.
\end{equation}
Note that the constant $K$  introduced in Sec.~\ref{generalDW} can be related to $\sigma$,
\begin{equation}
	K = \frac{\kappa^{2} \sigma}{2}.
\end{equation}

Next we compute the Ricci tensor $R^{\mu}{}_{\nu}$ of the metric \eqref{psmet}.
Similarly to the energy-momentum tensor, most of the components of $R^{\mu}{}_{\nu}$ vanish.
The non-vanishing components of the Ricci tensor are
\begin{align}
	&
	R^{0}{}_{0}
		= \frac{1}{2 A} \left( \frac{A''}{A} - \frac{A'^{2}}{A^{2}} + \frac{A' B'}{A B} \right),  \label{R00} \\
	&
	R^{1}{}_{1} = R^{2}{}_{2}
		= \frac{1}{2 A} \frac{B''}{B},  \label{R11} \\
	&
	R^{3}{}_{3}
		= \frac{1}{2 A} \left( \frac{A''}{A} - \frac{A'^{2}}{A^{2}}
			+ \frac{2 B''}{B} - \frac{B'^{2}}{B^{2}} - \frac{A' B'}{A B} \right).  \label{R33}
\end{align}
The $R^{1}{}_{1}$ and $R^{2}{}_{2}$ are the same because of the plane symmetry on $xy$-plane.
From eqs. \eqref{R00} - \eqref{R33} we find that the Ricci scalar $R$ is 
\begin{equation}  \label{Risca}
	R = \frac{1}{2 A}
		\left( \frac{2 A''}{A} - \frac{2 A'^{2}}{A^{2}} + \frac{4 B''}{B} - \frac{B'^{2}}{B^{2}} \right).
\end{equation}

Substituting the energy-momentum tensor and the Ricci tensor into the identity \eqref{identity}
we obtain the gravitational field equations.
In particular, $C_{0} - C_{1} = 0$ gives
\begin{equation}  \label{C01}
	\left( \frac{A''}{A} - \frac{A'^{2}}{A^{2}} - \frac{B''}{B} + \frac{A' B'}{A B} \right) f_{R}
		+ \left( \frac{A'}{A} - \frac{B'}{B} \right) f_{R}' = 0,
\end{equation}
and $C_{0} - C_{3} = 0$ gives,
\begin{equation}  \label{C03}
	\left( \frac{2 A' B'}{A B} - \frac{2 B''}{B} - \frac{B'^{2}}{B^{2}} \right) f_{R}
		+ \frac{2 A'}{A} f_{R}' - 2 f_{R}'' - 2 \kappa^{2} \phi'^{2} = 0.
\end{equation}
In addition we get from $C_{1} - C_{3} = 0$ that
\begin{equation}  \label{C13}
	\left( - \frac{A''}{A} + \frac{A'^{2}}{A^{2}} - \frac{B''}{B} + \frac{B'^{2}}{B^{2}} + \frac{A' B'}{A B} \right) f_{R}
		+ \left( \frac{A'}{A} + \frac{B'}{B} \right) f_{R}' - 2 f_{R}'' - 2 \kappa^{2} \phi'^{2} = 0.
\end{equation}
Although there are three gravitational field equations, in fact, one of them is derivable from another two.
Hence, there are in fact only two independent equations, the third one is redundant.

The full set of equations of motion and the gravitational equations cannot be solved exactly.
So, we are join got use some assumptions and approximations in next section, in order to find a solution.

\section{DW in $f(R)$ gravity}  \label{DWinf(R)}

In this section, we find an approximate solution to the differential equations obtained in the previous section.
First, we suppose that the corrections of $A$ and $B$ is small, so that
\begin{equation}  \label{ABcorrection}
	A(z) = 1 + a(z),
	\quad
	B(z) = 1 + b(z),
\end{equation}
where have assumed  $a$ and $ b$ of the order $\kappa^{2}$.
Under this approximation we have
\begin{equation}
	\frac{A'}{A} \approx a',
	\quad
	\frac{B'}{B} \approx b'.
\end{equation}
This approximation is the same as that used in \cite{Carames2011}.

Next, we expand $\phi$ along the lines of \cite{Widrow1989} as
\begin{equation}
	\phi = \phi_{0} + \kappa^{2} \sigma~,
\end{equation}
where $\mathcal{O}(\kappa^{4})$ is negligible.
Finally, as for the expression of $f_{R}(z)$, we assume 
\begin{equation}  \label{fR}
	f_{R}(z) = 1 + \psi_{0} z,
\end{equation}
where $\psi_{0}$ is a positive ``small" constant.
There is no special meaning in the expression of $f_{R}(z)$.
We have just taken it  for simplicity.
Given these approximations and assumptions,
the gravitational field equations \eqref{C01} - \eqref{C13} yield the following two equations:
\begin{align}
	&
	\alpha'' = 0,  \label{difeq1} \\
	&
	\beta'' + 2 \kappa^{2} \phi_{0}'^{2} = 0,  \label{difeq2}
\end{align}
where we have introduced
\begin{equation}
	\alpha'' = a'' - b'',
	\quad
	\beta'' = a'' + b''.
\end{equation}
We now solve these differential equation inside and outside of DW separately, in the case of a thick DW.
Recall that $\delta_{0}$ denotes the thickness of DW.
We find that the inside and the outside of DW can be represented 
by $z / \delta_{0} > 1$ and $z / \delta_{0} < 1$, respectively.
We make stronger conditions,
\begin{equation}  \label{in-out}
	\frac{z}{\delta_{0}} \gg 1,
	\quad
	\frac{z}{\delta_{0}} \ll 1,
\end{equation}
keeping all our assumptions and approximations.

Having obtained the correction terms $a, ~b$ for the metric $A, ~B$ from eqs. \eqref{difeq1} and \eqref{difeq2},
we calculate the corresponding $f(R)$ function.
First, let us  summarize our procedure.
We use the Ricci scalar $R$ obtained in eq.~\eqref{Risca}.
Applying the approximations of eq.~\eqref{ABcorrection} to $R$ we get
\begin{equation}  \label{Rapprox}
	R \approx \frac{1}{2} (3 \beta'' - \alpha''). 
\end{equation}
We can now calculate $R = R(z)$ outside and inside of DW by substituting $\alpha$ and $\beta$, respectively.
Next we can obtain $z = z(R)$.
By substituting $z$ into eq.~\eqref{fR}, we write down $f_{R}$ in terms of $R$.
Finally, we integrate that equation and derive the $f(R)$ functions.

\subsection{Outside DW}  \label{DWout}

First of all, we notice that eq.~\eqref{difeq1} implies
\begin{equation}  \label{alpha}
	\alpha = a - b = c_{1} z + c_{2},
\end{equation}
where $c_{1}, ~c_{2}$ are integration constants.
They can be determined by comparing eq.~\eqref{alpha} with static DW metric $A$ and $B$.
$A$ and $B$ are given by eq.~\eqref{stmet} outside of DW.
But we approximated $A$ and $B$ by eq.~\eqref{ABcorrection}.
Those approximated $A$ and $B$ should be equal to the expression in eq.~\eqref{stmet} being expanded in terms of $z$.
Therefore, $a$ and $b$ should be given by
\begin{equation}
	a = - \frac{1}{2} K z,
	\quad
	b = K z.
\end{equation}
Now we find that
\begin{equation}
	c_{1} = - \frac{3}{2} K,
	\quad
	c_{2} = 0,
\end{equation}
or
\begin{equation}  \label{alphaout}
	\alpha = - \frac{3}{2} K z.
\end{equation}

Next, we consider the expression of $\beta$.
Since $\phi_{0}$ is given by eq.~\eqref{phi}, substituting it into eq.~\eqref{difeq2} yields
\begin{equation}  \label{difeq3}
	\beta'' = - \frac{2 \kappa^{2} \eta^{2}}{\delta_{0}^{2}} {\rm sech}^{4} (e^{z / \delta_{0}}).
\end{equation}
The solution is
\begin{equation}  \label{betaout}
	\beta = - \frac{1}{2} K z
		+ \frac{4}{3} \kappa^{2} \eta^{2}
			\left[ \frac{z}{\delta_{0}} - \ln \left( e^{z / \delta_{0}} + e^{- z / \delta_{0}} \right)
				+ \left( e^{z / \delta_{0}} + e^{- z / \delta_{0}} \right)^{ -2} \right].
\end{equation}
We determined the values of the integration constants by evaluating $A, ~B$ in the small $z$ limit.

We are now in a position to derive the Ricci scalar $R$
by substituting eq.~\eqref{difeq1} and \eqref{difeq3} into eq.~\eqref{Rapprox},
\begin{equation}  \label{Rout}
	R = - \frac{3 \kappa^{2} \eta^{2}}{\delta_{0}^{2}} {\rm sech}^{4} (e^{z / \delta_{0}}).
\end{equation}
When $z / \delta_{0} \gg 1$, we can approximate ${\rm sech}^{4} (e^{z / \delta_{0}}) \approx 2^{4} e^{- 4 z / \delta_{0}}$.
Therefore, we obtain
\begin{equation}  \label{zout}
	z = \frac{\delta_{0}}{4} \ln \frac{2^{4} \cdot 3 \kappa^{2} \eta^{2}}{\delta_{0}^{2} (- R)}.
\end{equation}
Then eq.~\eqref{fR} yields
\begin{equation}  \label{fRout}
	f_{R}(R) = 1 + \frac{\psi_{0} \delta_{0}}{4} \ln \frac{2^{4} \cdot 3 \kappa^{2} \eta^{2}}{\delta_{0}^{2} (- R)}.
\end{equation}
Taking into account the stability condition, $f_{R} > 0$, we also get 
\begin{equation}
	- R < \frac{2^{4} \cdot 3 \kappa^{2} \eta^{2}}{\delta_{0}^{2}} e^{4 / \psi_{0} \delta_{0}} = R_{\rm max}.
\end{equation}
If $|R|$ is larger than $R_{\rm max}$, the system becomes unstable.
But the value $e^{4 / \psi_{0} \delta_{0}}$ is very large, since we have assumed earlier that $\psi_{0}$ is very small.
Therefore, the classical stability can be satisfied.
We can also calculate $f_{RR}(R)$ by differentiating $f_{R}(R)$ with respect to $R$,
\begin{equation}  \label{fRRout}
	f_{RR}(R) = \frac{\psi_{0} \delta_{0}}{4 (- R)}.
\end{equation}
This is positive definite, and, hence, the quantum stability is fulfilled too.

Thus the corresponding $f(R)$ function is given by
\begin{equation}  \label{fout}
	f(R) = \left( 1 - \frac{\psi_{0} \delta_{0}}{4} \right) R
		+ \frac{\psi_{0} \delta_{0}}{4} R \ln \frac{2^{4} \cdot 3 \kappa^{2} \eta^{2}}{\delta_{0}^{2} (- R)} + {\rm I. C.}
\end{equation}
We assume that $\psi_{0}$ is small. Then equation \eqref{fout} shows the Einstein-Hilbert action plus a small correction.
Though we have left the exponential term in the expression of $\beta''$, it is very small since $z / \delta_{0} \gg 1$.
Hence, if we take $e^{- 4 z / \delta_{0}} \approx 0$, then $\beta'' = 0$.
It leads to $R = 0$.
Therefore, our result is valid, since it is consistent with the known fact  that the space-time outside of DW is flat.
The integration constant in eq.~\eqref{fout} may be interpreted as a cosmological constant.

In the above derivation we merely considered the positions on a distance from DW.
How about the vicinity of DW?
To consider the environment near DW, we put
\begin{equation}
	\frac{z}{\delta_{0}} = 1 + \frac{\tilde{z}}{\delta_{0}},
\end{equation}
where $\tilde{z}$ is small.
Then eq.~\eqref{difeq2} becomes
\begin{equation}
	\beta'' = - \frac{2 e_{1} \kappa^{2} \eta^{2}}{\delta_{0}^{2}}
		\left[ \left( 1 + \frac{4 e_{2}}{\delta_{0}}  \right) - \frac{4 e_{2} z}{\delta_{0}} \right].
\end{equation}
where $e_{1}$ and $e_{2}$ are constants,
\begin{equation}
	e_{1} = \left( \frac{2 e}{e^{2} + 1} \right)^{4} \approx 0.18,
	\quad
	e_{2} = \frac{e^{2} - 1}{e^{2} + 1} \approx 0.76.
\end{equation}
It yields
\begin{equation}  \label{zvic}
	z = \left( 1 + \frac{1}{4 e_{2}} \right) \delta_{0} - \frac{\delta_{0}^{3} (- R)}{12 e_{1} e_{2} \kappa^{2} \eta^{2}}.
\end{equation}
Therefore, we have 
\begin{equation}  \label{fRvic}
	f_{R}(R) =
		\left( 1 + \psi_{0} \delta_{0} + \frac{\psi_{0} \delta_{0}}{4 e_{2}} \right)
			- \frac{\psi_{0} \delta_{0}^{3} (- R)}{12 e_{1} e_{2} \kappa^{2} \eta^{2}}.
\end{equation}
Taking into account that $\psi_{0}$ is small, $f_{R} > 0$ is satisfied.
Differentiating it with respect to $R$, we find that
\begin{equation}  \label{fRRvic}
	f_{RR}(R) = \frac{\psi_{0} \delta_{0}^{3}}{12 e_{1} e_{2} \kappa^{2} \eta^{2}}.
\end{equation}
is clearly positive.
Thus, indeed, the system under consideration is stable.
The $f(R)$ function can be derived as the Starobinsky-type expression $(f(R) \sim R + R^{2})$ \cite{Starobinsky1983} with
\begin{equation}  \label{fvic}
	f(R) = \left( 1 + \psi_{0} \delta_{0} + \frac{\psi_{0} \delta_{0}}{4 e_{2}} \right) R
		+ \frac{\psi_{0} \delta_{0}^{3}}{24 e_{1} e_{2} \kappa^{2} \eta^{2}} R^{2} + {\rm I. C.}
\end{equation}
The coefficients here are $4 e_{2} \approx 3.0$ and $24 e_{1} e_{2} \approx 3.3$.
If they are approximated by 3, we get 
\begin{equation}  \label{fvicap}
	f(R) = \left( 1 + \frac{4}{3} \psi_{0} \delta_{0} \right) R
		+ \frac{\psi_{0} \delta_{0}^{3}}{3 \kappa^{2} \eta^{2}} R^{2} + {\rm I. C.}
\end{equation}

\subsection{Inside DW}  \label{DWin}

In this subsection, we focus on the inside of DW.
In this case,  $\phi_{0}'^{2}$ can be expanded as
\footnote{
	The simplest option is to leave only the constant term in the expansion.
	However, it leads to almost the same result as in the case of the outside DW.
	Hence, we take the expansion up to the second order.
	}
\begin{equation}
	\phi_{0}'^{2} \approx \frac{\eta^{2}}{\delta_{0}^{2}} \left( 1 - \frac{z^{2}}{\delta_{0}^{2}} \right),
\end{equation}
Equation \eqref{difeq2} can be rewritten as
\begin{equation}  \label{difeq4}
	\beta'' = - \frac{2 \kappa^{2} \eta^{2}}{\delta_{0}^{2}} \left( 1 - \frac{z^{2}}{\delta_{0}^{2}} \right).
\end{equation}
Integrating it in terms of $z$ we obtain $\beta$
\begin{equation}  \label{betain}
	\beta = - \frac{2 \kappa^{2} \eta^{2}}{\delta_{0}}
		\left( \frac{1}{2} z^{2} - \frac{1}{12} \frac{z^{4}}{\delta_{0}^{2}} \right) + c_{3} z + c_{4},
\end{equation}
where $c_{3}$ and $c_{4}$ are the  integration constants.
The equation for $\alpha$ \eqref{difeq1} has the same form as that of the outside of DW.
Therefore, $\alpha$ can be written down in the form of eq.~\eqref{alpha},
\begin{equation}  \label{alphain}
	\alpha = c_{5} z + c_{6},
\end{equation}
with the two integration constants $c_{5}, ~c_{6}$.
However, the new integration constants we have now cannot be determined 
since the metric inside of DW is unknown.
But we can calculate the $f(R)$ function corresponding inside of DW.
For that purpose we need $\alpha''$ and $\beta''$, not $\alpha$ and $\beta$.

We can solve eq.~\eqref{difeq4} for $z$
\begin{equation}  \label{zin}
	z_{\pm} = \pm \delta_{0} \sqrt{1 - \frac{\delta_{0}^{2}}{3 \kappa^{2} \eta^{2}} (- R)}.
\end{equation}
The stability conditions
\begin{align}
	&
	f_{R}(R)
		= 1 \pm \psi_{0} \delta_{0} \sqrt{1 - \frac{\delta_{0}^{2}}{3 \kappa^{2} \eta^{2}} (- R)} > 0,  \label{fRin} \\
	&
	f_{RR}(R) = \pm \frac{\psi_{0} \delta_{0}}{6 \kappa^{2} \eta^{2}}
		\left( 1 - \frac{\delta_{0}^{2}}{3 \kappa^{2} \eta^{2}} (- R) \right)^{- 1 / 2} > 0,  \label{fRRin}
\end{align}
tell us that $z_{+}$ is valid but $z_{-}$ is not, while the constraint
\begin{equation}  \label{constraint}
	- R < \frac{3 \kappa^{2} \eta^{2}}{\delta_{0}^{2}}
\end{equation}
is required.
The corresponding $f(R)$ function is given by
\begin{equation}  \label{fin}
	f(R) = R + \frac{2 \psi_{0} \kappa^{2} \eta^{2}}{\delta_{0}}
		\left( 1 - \frac{\delta_{0}^{2}}{3 \kappa^{2} \eta^{2}} (- R) \right)^{3 / 2} + {\rm I. C.}
\end{equation}
The constraint \eqref{constraint} is satisfied when $R \approx 0$.
If DW exists in our universe, it is known that $\eta \lessim 100 ~[\rm keV]$  \cite{Friedland2003}.
Hence, the result we obtained may apply to the DW at present.

There are also the models which require a large $\eta$ value.
One of such models is the model of topological inflation \cite{Linde1994a}, \cite{Vilenkin1994},
which is similar to chaotic inflation,
but with the topological defects created in the early universe which are exponentially expanding.
Those defects in the topological inflation may continue to expand somewhere in our universe.
The required value of $\eta$ should be $\eta \sim \kappa^{- 1}$
in this type of inflation \cite{Linde1994a, Vilenkin1994, Sakai1996}.
For this $\eta$, the stability conditions corresponding to our model \eqref{fin} can be satisfied.
Then, if $\lambda \sim 1$, the right hand side of eq.~\eqref{constraint} is of the order $\kappa^{- 2}$ and,
therefore, eq.~\eqref{constraint} can be $- R \ll 3 \kappa^{2} \eta^{2} / \delta_{0}^{2}$.
Hence, the second term of eq.~\eqref{fin} can be expanded as
\begin{equation}  \label{finex}
	f(R) = \frac{2 \psi_{0} \kappa^{2} \eta^{2}}{\delta_{0}} + (1 + \psi_{0} \delta_{0}) R,
\end{equation}
where the integration constant is absorbed into first term.

\subsection{Estimate for $\psi_{0}$}

Of course, it is desirable to get the observational bound for our parameter $\psi_{0}$.
According to CODATA \cite{Mohr2012}, the value of the Newton's constant $G_{\rm N}$ is given by
\begin{equation}
	G_{\rm N} = 6.6738(80) \times 10^{- 11}~[\rm m^{3} / kg \cdot s^{2}],
\end{equation}
in the SI units.
The solution outside DW \eqref{fout} should duplicate the Einstein-Hilbert action in the low-curvature (small $|R|$) regime.
When $|R| \ll 1$, the logarithm term can be dropped, so that
\begin{equation}  \label{lowfout}
	f(R) \approx - \left( 1 - \frac{\psi_{0} \delta_{0}}{4} \right) R.
\end{equation}
Equation \eqref{lowfout} means that the correction $\psi_{0} \delta_{0}$ is added to the Newton's constant.
This correction should be smaller than $10^{- 4}$ in order to be undetectable.
Therefore, we obtain the constraint, $\psi_{0} \delta_{0} \lessim 10^{- 4}$.
For instance, if $\lambda \sim 1$, we have
\begin{equation}
	\psi_{0} \lessim 10^{- 4} \eta.
\end{equation}
Hence, by using  the present DW constraint $\eta \lessim 100~[\rm keV]$,
we derive $\psi_{0} \lessim 10~[\rm eV] \sim 10^{8}~[\rm 1 / m]$.
In the case of the topological inflation $\eta \sim \kappa^{- 1} \sim 10^{19}~[\rm GeV]$,
we get $\psi_{0} \lessim 10^{15}~[\rm GeV] \sim 10^{31}~[\rm 1 / m]$.

Of course, we get only the upper bounds for $\psi_{0}$.
If it takes a smaller value, the correction to the gravitational constant $G_{\rm N}$ will appear in the next digits.
Therefore, we conclude that the value of $\psi_{0}$ needs to be measured more precisely.

\section{Conclusion}  \label{Conc}

In this paper we derived the gravitational field inside and outside of DW in terms of the $f(R)$ functions.
The gravitational field equations in the $f(R)$ gravity are more complicated than those in General Relativity,
and they cannot be exactly solved.
Our results are derived in the weak field approximation.
Applying the stability conditions, we have chosen the adequate solutions.
The solution outside DW is given by eq.~\eqref{fout}, and in the vicinity of DW is given by eq.~\eqref{fvic}.
Inside DW we got eq.~\eqref{fin}.

Our results satisfy the constraint to exist DW in the present universe.
In addition, those can be applied to the models which require large $\eta$ value, like the topological inflation.
In deriving our solutions, we have introduced the constant $\psi_{0}$
which has the inverse length dimension and yields a correction to the Einstein gravity.
We determined the upper bound on the absolute value of $\psi_{0}$
from the known value of the gravitational constant $G_{\rm N}$.
The magnitude of this correction is thus very small.
Therefore, more precise observations are necessary to detect it.
Though $\psi_{0}$ is likely to correspond to the constant $K$ present in the static plane symmetric metric,
their precise relation still remains to be found.

Our solution of inside DW should be useful for describing the topological inflation in detail.
However, we assumed the DW to be static in the present paper.
In contrast, the DW in the topological inflation continues to expand.
Therefore, we should take into account the time evolution of DW, which we hope to consider elsewhere.

\section*{Acknowledgements}

I would like to express my deep gratitude to S. V. Ketov for useful discussions.

\end{document}